\documentclass[conference]{IEEEtran}
\IEEEoverridecommandlockouts
\usepackage{cite}
\usepackage{amsmath,amssymb,amsfonts}
\usepackage{algorithmic}
\usepackage{graphicx}
\usepackage{textcomp}
\usepackage{subcaption}
\usepackage{multirow}
\usepackage{adjustbox}
\usepackage{array}
\usepackage{xcolor}
\usepackage{float}
\usepackage[left=0.680in,right=0.633in,top=0.764in,bottom=1.049in]{geometry}
\def\BibTeX{{\rm B\kern-.05em{\sc i\kern-.025em b}\kern-.08em
    T\kern-.1667em\lower.7ex\hbox{E}\kern-.125emX}}
\begin{document}
\title{Integrated Space Domain Awareness and Communication System}
\author{
\IEEEauthorblockN{Selen Gecgel Cetin}
\IEEEauthorblockA{\textit{Department of Electronics}\\
                  \textit{and Communication Engineering}\\
                  \textit{Istanbul Technical University}\\Istanbul, Turkey\\
                  {gecgel16@itu.edu.tr}
                  }
\and
\IEEEauthorblockN{Berna Ozbek}
\IEEEauthorblockA{\textit{Department of Electrical}\\
                  \textit{and Electronics Engineering}\\
                  \textit{Izmir Institute of Technology}\\Izmir, Turkey\\
                  {bernaozbek@iyte.edu.tr}
                  }
\and                  
\IEEEauthorblockN{Gunes Karabulut Kurt}
\IEEEauthorblockA{\textit{Department of Electrical Engineering}\\
                  \textit{Polytechnique Montreal}\\Montreal, QC, Canada\\
                  {gunes.kurt@polymtl.ca}
                  }
}
\maketitle
\begin{abstract}
Space has been reforming and this evolution brings new threats that, together with technological developments and malicious intent, can pose a major challenge. Space domain awareness (SDA), a new conceptual idea, has come to the forefront. It aims sensing, detection, identification and countermeasures by providing autonomy, intelligence and flexibility against potential threats in space. In this study, we first present an insightful and clear view of the new space. Secondly, we propose an integrated SDA and communication (ISDAC) system for attacker detection. We assume that the attacker has advanced communication capabilities to vary attack scenarios, such as random attacks on some receiver antennas. To track random patterns and meet SDA requirements, a lightweight convolutional neural network architecture is developed. The proposed ISDAC system shows superior and robust performance under 12 different super-attacker configurations with a detection accuracy of over $97.8\%$.
\end{abstract}
\section{Introduction}
As we move toward the sixth generation (6G), wireless networks will take on a new dimension with the inclusion of satellite communications, but they will also put a different complexion on space, which is being called "\textit{New Space}" \cite{ref1}. Satellite communications, which enable the provision of services at affordable costs, can provide adequate service, especially in suburban and rural areas that may not be served by terrestrial networks. LeoSat, TeleSat, Honyan and O3b are taking their place in the new space, while well-known pioneers OneWeb and Starlink have already set out to increase the magnitude of their constellations. Satellite communications not only bring numerous benefits, but also significant security requirements and issues that are becoming more diverse as a consequence of these developments in new space \cite{ref2}.
\begin{figure}[!ht]
  \center
  \includegraphics[width=1\linewidth]{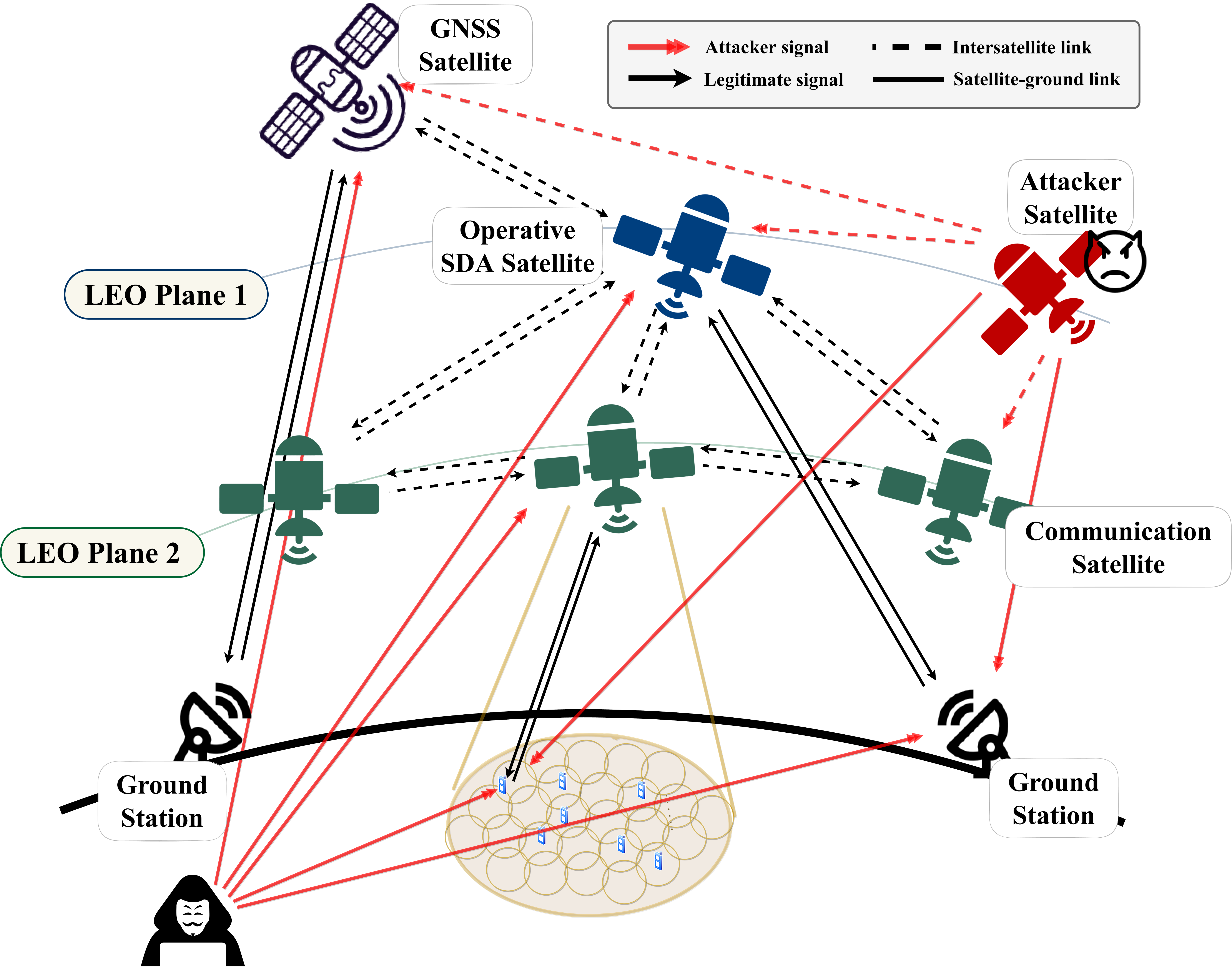}
  \caption{A glimpse of new space from a security perspective, considering three key satellites with different missions that can be threatened by an attacker satellite. The satellites in LEO plane 1 provide communications services to Earth, while a GNSS satellite provides PNT information to ground and space systems. The operative SDA satellite plays the role of managing and supporting all SDA applications within the constellation in LEO plane 1.}
  \label{fig1}
  \vspace{-0.5cm}
\end{figure}

Satellite communications security has several signal vulnerabilities and is susceptible to attacks such as jamming, spoofing, eavesdropping, and signal injection. Over the past decade, these vulnerabilities have become apparent in research studies, but mostly through intentional/unintentional attacks \cite{pavurthesis_ref20,pavurthesis_ref83,pavurthesis_ref37,pavurthesis_ref411,pavurthesis_ref417}. However, we first became aware of the looming threat from China's complaint noted that the Chinese space station had encountered Starlink satellites twice in 2021. The complaint specifically highlighted maneuvering difficulties during the second incident due to the unpredictable movement of the Starlink satellite. In the wake of Russian jamming attempts on Starlink in Ukraine, both commercial and national actors in space are aware of the hazard. Current satellites are obviously vulnerable to threats and attackers that can have more devastating and irretrievable consequences, like collisions \cite{me1}. In addition, the sources of potential threats to satellites are not only ground-based, but may also be in space, as shown in Figure \ref{fig1}. 

If satellites play a role in critical areas, their security must be examined more closely. Recently, a new concept has come to the fore; \textit{space domain awareness} (SDA). It defines the comprehension of the operational space environment, threats and vulnerabilities surrounding functional areas such as satellite communications or positioning, navigation and timing (PNT). SDA applications aim to support the better decision-making in all functional areas. Therefore, new solutions for security must be nested with SDA applications \cite{pavur_survey, sda5, sda4, Falco, sda1}. In the light of all the developments and advances, and taking into account the concerns in space, we have proposed an integrated SDA and communication (ISDAC) system that is leveraged with detection mechanisms based on machine learning.
\subsection{Challenges and Motivations}
We first explained the main challenges and motivations that inspired our view of the new space and originated contributions in this paper. The main challenges that also need to be certainly considered in further research studies are listed below:
\begin{itemize}
    \item[1.] The nature and impact of attacks on the satellite mission depend on various conditions, such as the transmission path, the relative speed of the attacked satellite, signal propagation, environmental conditions, and the position of the legitimate entity. These observations are also critical to the security solutions that are usually performed at the ground station in current systems. This can make the solutions weaker against attacks from space.
    \item[2.] Security solutions must meet the stringent requirements of new space missions, such as autonomy, flexibility, and compatibility with instantaneous or exceptional conditions. In particular, meeting these requirements with no or less human control seems to be the most important need, since \textit{timely intervention} is not as effortless in space as it is in terrestrial or aerial networks.  
    \item[3.] Resources in space are finite, and spacecraft and equipment are computationally constrained. These challenges involve trade offs in performance, sustainability, and security.
    \item[4.] If satellite communications are to play a role in the next generations, they must perform the key essentials such as high data rate and spectrum efficiency.
    \item[5.] New technologies and techniques equip attackers with enhanced communication and computational capabilities. Attacks can be easily carried out with small and affordable devices such as software-defined radios, single-board computers, or hobbyist devices. In addition, these devices are easy to program and configure for various attack scenarios.
\end{itemize}
In the above challenges, there are many sub-issues that need to be considered in terms of system aspects, applications, and use cases. However, they are not an excuse to let up progress for the new space security. Below we have listed our motivations for the proposed ISDAC system:
\begin{itemize}
    \item The important part of current satellites are equipped with cold-war technologies and systems that suffer from security vulnerabilities.
    \item The major prerequisites of SDA for further systems are autonomy, intelligence, and robustness for SDA's critical milestones: sensing, detection, identification, and countermeasure. 
    \item Machine learning algorithms, thanks to their ability to heuristically learn and flexibly improve, enable better decision-making systems and are therefore found in cutting-edge technologies.
\end{itemize}
\subsection{Contributions}
We have listed our contributions below, taking into account all the above considerations:
\begin{itemize}
    \item We present our view of the new space, which highlights potential threats and provides insights into future solutions based on SDA, as shown in Figure \ref{fig1}. In this regard, we propose an operational SDA satellite to support the management and control of satellites rather than just  managing them on the ground (considering the first challenge).
    \item We propose a machine learning driven SDA application to detect the competent attacker who can randomize or change the attacks (considering the second and fifth challenges).
    \item Our the intelligent SDA application is integrated with the existing satellite communication system. In this way, our solution also eliminates the need for an additional sensing system (considering the third challenge).
    \item We have developed a lightweight convolutional neural network (CNN) architecture and used tensor-based data generation, which is less complex compared to image-based approaches commonly used in the literature (considering the third challenge).
    \item We consider a spectrally efficient multi-antenna system and examined our solution in 12 different cases where the attacker is equipped with advanced capabilities. All cases consist of random attacks on some receive antennas, changes in attack strategies, and variations in the power level of the attacker and/or the legitimate transmitter. The proposed solution shows a competent and consistent detection performance in all cases and achieves over $97.8\%$ accuracy against the super-attacker (considering the fourth and fifth challenges).
\end{itemize}
\section{Related Works}
SDA applications in current satellite systems are based on age-old practices and there are several burning issues related to different aspects. However, there is an almost complete lack of research studies in the literature. In \cite{sda5}, the authors aim a SDA application that provides insights for orbit determination by sensing and identifying objects in space. Researchers at the Ottawa Research Lab adapt an existing antenna for an SDA sensing application and achieve an accurate measurement of the one-way Doppler and perform interferometry by using the two antennas \cite{sda4}. 
In \cite{Falco}, an SDA detection application is presented by considering malicious nature of radio frequency interference against Global Navigation Satellite System (GNSS) signals. The authors reformat the received signal as an image to train machine learning algorithms and achieve over $99\%$ accuracy. In \cite{sda1}, the authors present a knowledge integration framework that collects data from various sensors and resources such as texts, and news, and reports. They also propose an anomaly detection mechanism for SDA using this framework.

There are numerous satisfactory studies in the literature that present different techniques for jamming attacks on satellite communications \cite{jam5}. However, considering the new space and communication system advances holistically with SDA requirements, research studies are limited. Most studies focus on the satellite systems that will be fallen into obscurity in the foreseeable future. Another large portion of existing studies only examine ground-based jamming attacks. Another missing point is that jamming attacks are generally analyzed against satellites in geosynchronous Earth orbit (GEO) \cite{jam2}. However, the new space is expected to be teeming with satellites in low Earth orbit (LEO), and there are important differences between GEO and LEO satellites such as speed, distance, and delay. Moreover, the conventional methods such as hypothesis testing and rule-based methods may fail against the random behaviors of the attackers because they are mostly based on assumptions.
\section{System Description}
\begin{figure}[!t]
  \center
  \includegraphics[width=0.85\linewidth]{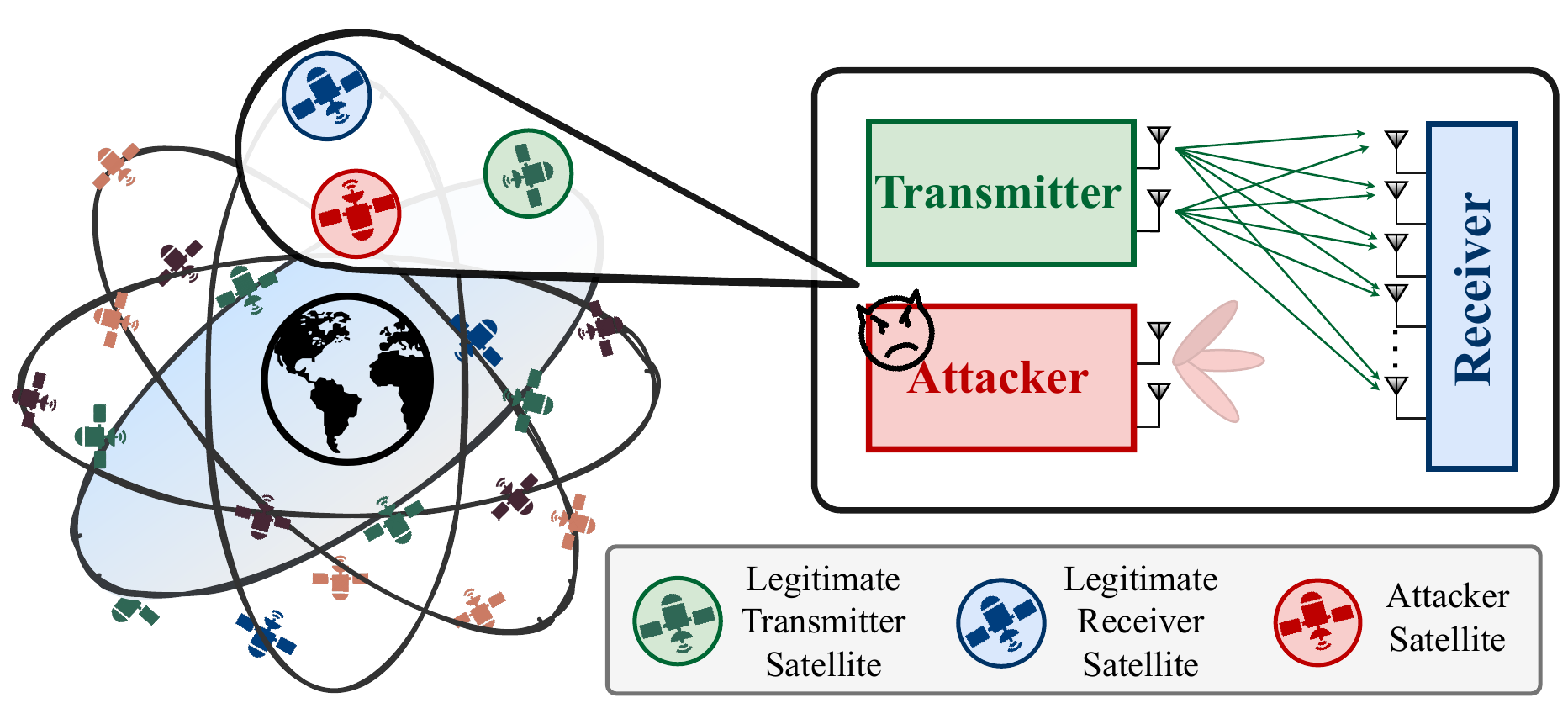}
  \caption{The overview of MIMO-OFDM based legitimate and attacker satellite systems where the attacker has beam-steering antennas and randomly targets the some antennas of the receiver.}
  \label{fig2}
  \vspace{-0.5cm}
\end{figure}
Satellite communications, especially based on LEO satellites, have become a component of the next generation broadband and wireless systems and will undoubtedly occupy a large place in the next generations. They must meet the key requirements of future systems such as high data rates and bandwidth efficiency. Multiple-input multiple-output (MIMO) systems can increase spectral efficiency or SNR, which is an important phenomenon due to higher propagation losses in space thanks to diversity benefits. On the other hand, orthogonal frequency division multiplexing (OFDM) offers flexibility in spectrum management and is well suited for integration with MIMO systems when the problem of interference between antennas is taken into account \cite{OFDM1, Rician}. In addition, OFDM allows the dimensioning of features in the data by forming the signal in the frequency domain, i.e., it naturally helps to improve detection performance. Therefore, a MIMO-OFDM based satellite communication is considered in this study.

The legitimate system has $N_t$ transmit antennas and $N_r$ receive antennas. The data is modulated with binary phase-shift keying modulation at each transmit antenna. To estimate channel state information, pilot symbols are inserted uniformly between subcarriers based on comb-type pilot arrangement as
\begin{equation}
    X_t(k) = \left\{\begin{matrix}
X_{tp}(m) \ , & \bmod{(k,d)} = 4, \\ 
0, & \ \ \text{otherwise} 
\end{matrix}\right. 
\end{equation}
where $X_t(k)$ denotes $k^{th}$ subcarrier of an OFDM symbol at $t^{th}$ transmit antenna and $k=0,1,\cdots,N-1$ where $N$ is the number of subcarriers. $d$ is the interval between consecutive pilot symbols and $m = 0, 1, \cdots ,(N_p-1)$ where ${N_p}$ is the number of pilot symbols. The data sequence in frequency is converted to time-domain signals by inverse fast Fourier transform and guard time interval (cyclic prefix) is appended. Rician channel model is employed by taking into account the line-of-sight propagation \cite{Rician}.

The received signal at $r^{th}$ antenna is represented in frequency domain as follows:
\begin{equation}
    \mathbf{Y}_{r} = \sum_{t=1}^{N_{t}} \mathbf{X}_{t} \mathbf{H}^{l}_{tr} + \mathbf{W}_{r},
\end{equation}
where $\mathbf{Y}_r =\bigl[Y_r(0), \ Y_r(1),\ \cdots, \ Y_r(N-1)\bigl]^{T}$, $\mathbf{X}_t = \mathrm{diag}\left \{ \bigl[X_t(0),\ X_t(1),\ \cdots,\ X_t(N-1)\bigl] \right \}$ is transmitted symbol sequences from $t^{th}$ antenna, $\mathbf{W}_r = \bigl[W_r(0),\ W_r(1),\ \cdots,\ W_r(N-1)\bigl]^{T}$ is zero mean white Gaussian vector and the superscript $\left [ . \right ]^T$ indicates transpose. $\mathbf{H}^{l}_{tr} = \bigl[H^{l}_{tr}(0),\ H^{l}_{tr}(1),\ \cdots,\ H^{l}_{tr}(N-1)\bigl]^{T}$ is frequency response of the channel between $t^{th}$ transmit antenna and $r^{th}$ receive antenna of legitimate satellites.
\subsection{Super-Attacker Model}
We assume that the super-attacker is equipped with multiple antennas that can be steered to transmit signals in a specific direction and it is capable of jamming in two different ways: pilot tone jamming and barrage jamming. The received signal under one of these jamming attacks is defined as
\begin{equation}\label{attackedsignal}
    \mathbf{Y}_{r} = \sum_{t=1}^{N_{t}} \mathbf{X}_{t} \mathbf{H}^{l}_{tr} + \mathbf{W}_{r} + {\gamma}_r \mathbf{J}_{t} \mathbf{H}^{a}_{tr}
\end{equation}
where $\mathbf{J}_t = \mathrm{diag} \left \{ \bigl[J_t(0),\ J_t(1), \cdots,\ J_t(N-1)\bigl] \right \}$ is the jamming signal vector in frequency domain and ${\gamma}_r \in \left \{ 0, 1 \right \}$ is attack indicator. ${\gamma}_r = 1$ if the $r^{th}$ receive antenna is attacked, otherwise equal to zero. $\mathbf{H}^{a}_{tr} = \bigl[H^{a}_{tr}(0),\ H^{a}_{tr}(1),\ \cdots,\ H^{a}_{tr}(N-1)\bigl]^{T}$ is channel in frequency domain between $t^{th}$ transmit antenna of the attacker satellite and $r^{th}$ receive antenna of the legitimate satellite.
\begin{figure}[!ht]
\center
\includegraphics[width=0.95\linewidth]{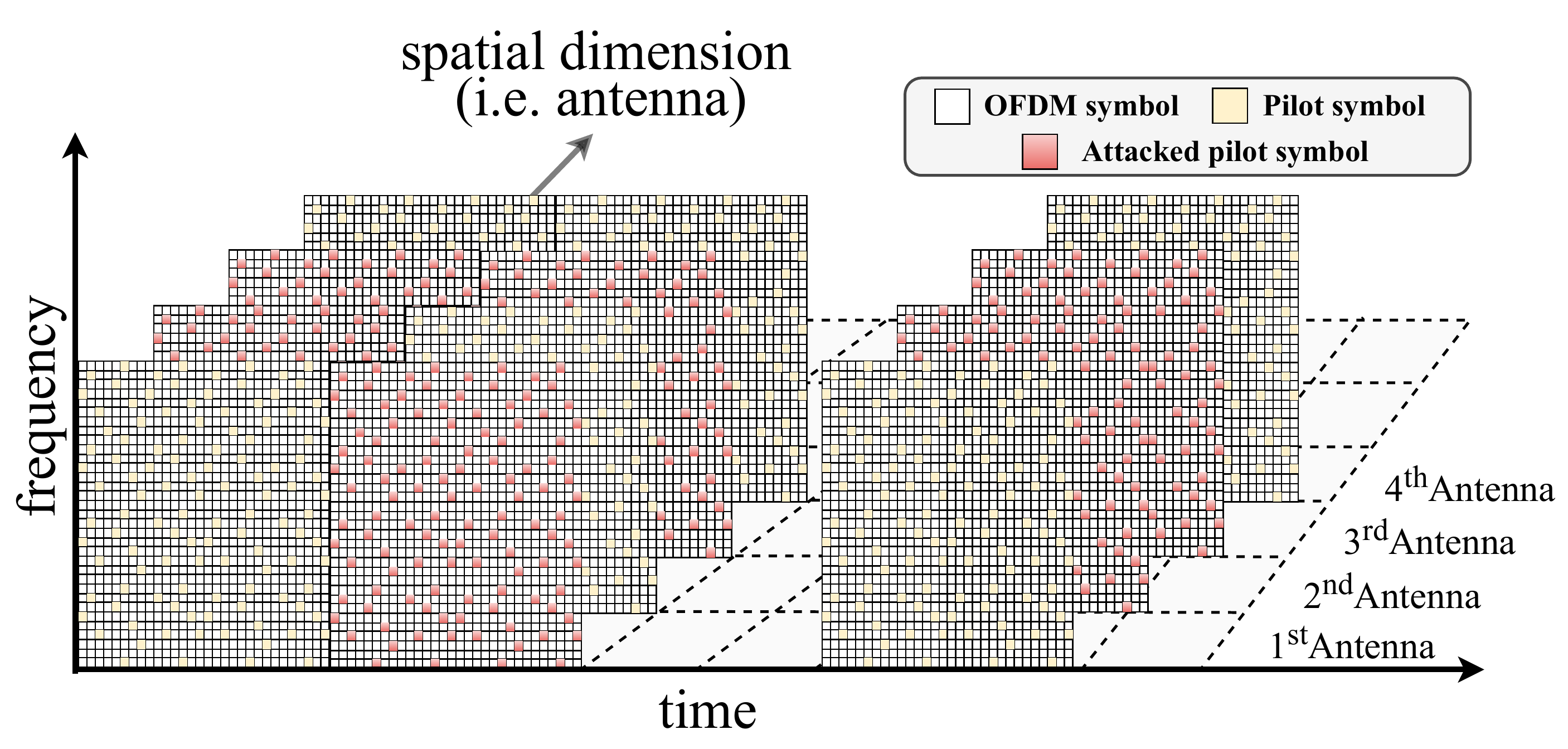}
\caption{A representation of the spectrum where two antennas of the legitimate satellite receiver are attacked at different time intervals by the super attacker using pilot tone jamming technique  without the constraint of time overlap.}
\label{fig3}
\vspace{-0.5cm}
\end{figure}
\subsubsection{Barrage Jamming}
When the target signal is not known in advance, barrage jamming is an optimal jamming attack. It is simply defined as a noise attack on the entire transmission bandwidth \cite{barrage}. In this type of attack, the received signal in (\ref{attackedsignal}) is defined by the jamming signal vector $\mathbf{J_t}$, which is a white Gaussian noise with distribution of $\mathcal{CN}(0,\,{\sigma}^{2})$.
\subsubsection{Pilot Tone Jamming}
Only the frequencies of the pilot tones are subject to jamming attacks; therefore, it depends on knowledge of the positions of the pilot symbols. This makes the pilot tone jamming attack power-efficient compared to the barrage jamming attack \cite{pilot}.
\section{Space Domain Awareness Application}
SDA applications can be categorized over four main areas: sensing, detection, identification, and countermeasure for the surrounding issues of a satellite. However, as we mentioned in the first sections, satellites have limited computational capabilities and are constrained in terms of certain resources such as energy. This results in the requirement that SDA solutions should be intertwined with available satellite systems when they meet the demand of autonomy, intelligence, and competence. Therefore, our goal is to achieve intelligent SDA detection by integrating satellite communication systems and eliminating the need for sensing.

Our proposed SDA application for attacker detection is based on a CNN that is highly eligible to comprehend a 3-dimensional (3D) tensor due to its convolutional filtering property. The proposed CNN architecture consists of 2 convolutional layers, a flatten layer and 2 fully-connected layers. The convolutional layers have 32 and 16 neurons with $3 \times 3$ kernels, while the first fully-connected layer has 96 neurons. Initialization of all kernels is randomly generated with a uniform distribution. Batch normalization and dropout layers (at a rate of 0.5) are used to avoid overfitting during the training phase. The ReLU activation function is used for simplification in all layers except the last layer where the softmax activation function is used. The adaptive moment estimation algorithm is chosen with a learning rate of 0.01, ${\beta}_1$ 0.9 and ${\beta}_2$ 0.99 for optimization
. The designed CNN architecture is optimized by using the categorical-cross entropy loss function.
\section{Data Acquisition}
A communication signal is generated using the above MIMO-OFDM and attacker systems with the parameters set in the Table \ref{params}. The received signals at all antennas are recorded over a time interval equal to 10 frames and plotted as 2D tensors. The 2D tensors can be seen in the Figure \ref{fig3} where the two receiver antennas are randomly targeted by a pilot tone jammer. Short time-frequency transform and a mean filter are applied to the 2D tensors to smooth the data and reduce fluctuations. This also reduces the computational overhead. 2D tensors are stacked in the third dimension, which represents the spatial domain. The composed 3D tensor represents a data sample.

\begin{table}[!h]
\center
\caption{The parameter setting for the generation of the communication signals.}
\label{params}
\resizebox{0.5\linewidth}{!}{%
\begin{tabular}{|l|c|l}
\cline{1-2}
\multicolumn{1}{|c|}{\textbf{Parameter Name}} & \textbf{Value} &  \\ \cline{1-2}
Guard time interval                           & 64             &  \\ \cline{1-2}
\# of frames in a transmitted signal          & 10             &  \\ \cline{1-2}
\# of symbols in a frame                      & 60             &  \\ \cline{1-2}
\# of subcarriers                             & 1024           &  \\ \cline{1-2}
\# of data subcarriers                        & 705            &  \\ \cline{1-2}
$N_p$                                         & 88             &  \\ \cline{1-2}
$d$                                           & 8              &  \\ \cline{1-2}
$N_t$                                         & 2              &  \\ \cline{1-2}
$N_r$                                         & 4, 8           &  \\ \cline{1-2}
Rician channel K factor                       & 5              &  \\ \cline{1-2}
\end{tabular}%
}
\end{table}
In this study, 12 different data sets are created by following aforementioned steps. Each data set is formed to analyze the proposed CNN based detection system under different combinations of critical considerations: attack type, SJR and SNR levels, different receive antenna numbers with the same number of attacked antennas. The data sets are categorized with 4 groups: A, B, C, D, which are enumerated as shown in the Table \ref{table}. Both the attacker and the legitimate satellite have 2 antennas on the transmitter side during the generation of all data sets. The legitimate receiver satellite has 4 antennas for the data sets of groups A and C, and 8 antennas for the data sets of groups B and D. Groups A and B each have 5 sub-datasets consisting of 150 data samples for each SNR, all SJRs, and both jamming attacks, for a total of 3000 samples. In contrast, groups C and D have a data set consisting of 7500 samples formed under all considerations.
\begin{table}[!h]
\center
\caption{The detailed information about data configurations and the number of data samples in all data sets.}
\label{table}
\begin{adjustbox}{max width=0.95\columnwidth}
{
                        \begin{tabular}{|ccc|ccccccl|c}
                        \cline{1-10}
                        \multicolumn{3}{|c|}{\textbf{Group Identity}}                                                                                                                                                                                                                 & \multicolumn{5}{c|}{\text{A and B}}                                                                                                                                   & \multicolumn{2}{c|}{\text{C and D}}      & \textbf{} \\ \cline{1-10}
                        \multicolumn{3}{|c|}{\textbf{Data set Identity Number}}                                                                                                                                                                                                               & \multicolumn{1}{c|}{\text{1}} & \multicolumn{1}{c|}{\text{2}} & \multicolumn{1}{c|}{\text{3}} & \multicolumn{1}{c|}{\text{4}} & \multicolumn{1}{c|}{\text{5}} & \multicolumn{2}{c|}{\text{1}}            &           \\ \cline{1-10}
                        \multicolumn{1}{|c|}{\multirow{2}{*}{\textbf{Attacker}}} & \multicolumn{1}{c|}{\multirow{2}{*}{\textbf{Jamming}}}                                                        & \multirow{2}{*}{\textbf{\begin{tabular}[c]{@{}c@{}}SJR (dB)\end{tabular}}} & \multicolumn{7}{c|}{\textbf{SNR (dB)}}                                                                                                                                                                               &           \\ \cline{4-10}
                        \multicolumn{1}{|c|}{}                                   & \multicolumn{1}{c|}{}                                                                                         &                                                                              & \multicolumn{1}{c|}{5}          & \multicolumn{1}{c|}{10}         & \multicolumn{1}{c|}{15}         & \multicolumn{1}{c|}{20}         & \multicolumn{1}{c|}{25}         & \multicolumn{2}{c|}{5, 10, 15, 20, 25}     &           \\ \cline{1-10}
                        \multicolumn{1}{|c|}{\multirow{10}{*}{\textbf{Present}}} & \multicolumn{1}{c|}{\multirow{5}{*}{\textbf{\begin{tabular}[c]{@{}c@{}}Barrage\\ Jamming\end{tabular}}}}      & 0                                                                            & \multicolumn{5}{c|}{\multirow{10}{*}{150}}                                                                                                                              & \multicolumn{2}{c|}{\multirow{10}{*}{375}} &           \\ \cline{3-3}
                        \multicolumn{1}{|c|}{}                                   & \multicolumn{1}{c|}{}                                                                                         & -5                                                                           & \multicolumn{5}{c|}{}                                                                                                                                                   & \multicolumn{2}{c|}{}                      &           \\ \cline{3-3}
                        \multicolumn{1}{|c|}{}                                   & \multicolumn{1}{c|}{}                                                                                         & -10                                                                          & \multicolumn{5}{c|}{}                                                                                                                                                   & \multicolumn{2}{c|}{}                      &           \\ \cline{3-3}
                        \multicolumn{1}{|c|}{}                                   & \multicolumn{1}{c|}{}                                                                                         & -15                                                                          & \multicolumn{5}{c|}{}                                                                                                                                                   & \multicolumn{2}{c|}{}                      &           \\ \cline{3-3}
                        \multicolumn{1}{|c|}{}                                   & \multicolumn{1}{c|}{}                                                                                         & -20                                                                          & \multicolumn{5}{c|}{}                                                                                                                                                   & \multicolumn{2}{c|}{}                      &           \\ \cline{2-3}
                        \multicolumn{1}{|c|}{}                                   & \multicolumn{1}{c|}{\multirow{5}{*}{\textbf{\begin{tabular}[c]{@{}c@{}}Pilot\\ Tone\\ Jamming\end{tabular}}}} & 0                                                                            & \multicolumn{5}{c|}{}                                                                                                                                                   & \multicolumn{2}{c|}{}                      &           \\ \cline{3-3}
                        \multicolumn{1}{|c|}{}                                   & \multicolumn{1}{c|}{}                                                                                         & -5                                                                           & \multicolumn{5}{c|}{}                                                                                                                                                   & \multicolumn{2}{c|}{}                      &           \\ \cline{3-3}
                        \multicolumn{1}{|c|}{}                                   & \multicolumn{1}{c|}{}                                                                                         & -10                                                                          & \multicolumn{5}{c|}{}                                                                                                                                                   & \multicolumn{2}{c|}{}                      &           \\ \cline{3-3}
                        \multicolumn{1}{|c|}{}                                   & \multicolumn{1}{c|}{}                                                                                         & -15                                                                          & \multicolumn{5}{c|}{}                                                                                                                                                   & \multicolumn{2}{c|}{}                      &           \\ \cline{3-3}
                        \multicolumn{1}{|c|}{}                                   & \multicolumn{1}{c|}{}                                                                                         & -20                                                                          & \multicolumn{5}{c|}{}                                                                                                                                                   & \multicolumn{2}{c|}{}                      &           \\ \cline{1-10}
                        \multicolumn{1}{|c|}{\textbf{Absent}}                    & \multicolumn{1}{c|}{N/A}                                                                                      & N/A                                                                          & \multicolumn{5}{c|}{1500}                                                                                                                                               & \multicolumn{2}{c|}{3750}                  &           \\ \cline{1-10}
                        \multicolumn{3}{|c|}{\textbf{Total Number of Samples}}                                                                                                                                                                                                  & \multicolumn{5}{c|}{3000}                                                                                                                                               & \multicolumn{2}{c|}{7500}                  &           \\ \cline{1-10}
                        \end{tabular}
}
\end{adjustbox}
\end{table}
\vspace{-0.2cm}
\section{Results}
\begin{figure}[!ht]
  \center
  \includegraphics[width=0.8\linewidth]{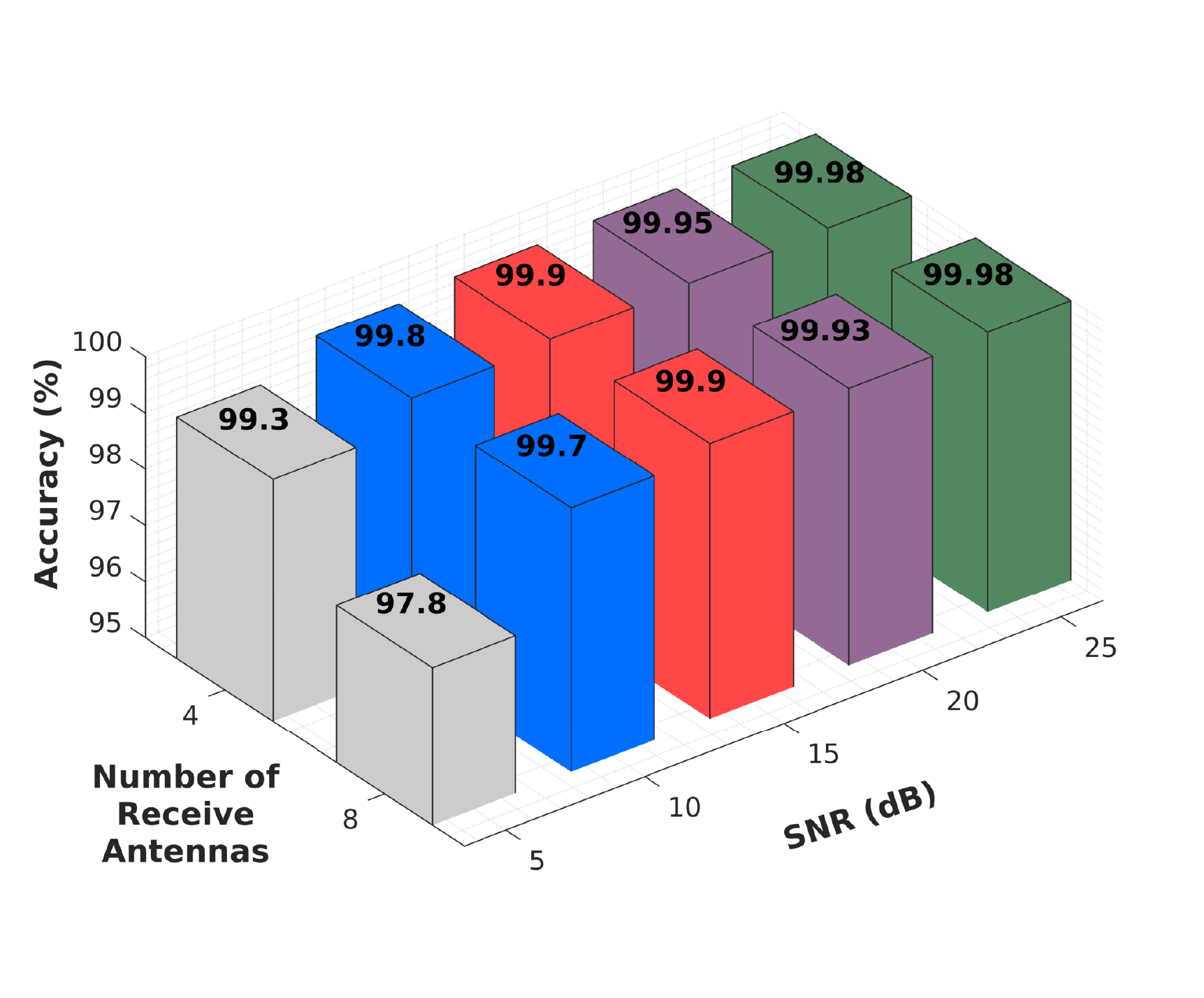}
  \caption{Performance comparison of proposed intelligent SDA detection at each SNR and different number of receive antennas.}
  \label{fig4}
  \vspace{-0.5cm}
\end{figure}
The performance of the proposed intelligent SDA detection is investigated with 12 data sets, i.e., 12 attack scenarios under different system settings. The main goal for each scenario/case is to determine the absence or presence of the attacker. The designed CNN architecture is trained and tested with the same hyperparameter settings by using an offline learning mechanism. All group A and B data sets are divided with the rate of $60\%$ and $40\%$ for training and testing phase, respectively, while group C and D data sets are split equally.

The obtained results are presented in Figure \ref{fig4} and Table \ref{table3} in terms of accuracy. The results are above $97.8\%$ accuracy, which proves the robustness of the proposed SDA detection to different jamming power or attacker strategies, since all data sets were generated considering that the SJR may exhibit variations or the type of jamming may be changed by the attacker. The results also point to the durability and flexibility of machine learning-based solutions in the face of advanced attacks.

The results in Figure \ref{fig4} show that increasing the number of antennas at the receiver while keeping the number of attacked antennas constant leads to a slight degradation in performance. The difference in detection performance becomes significant at the lowest SNR, reaching $1.5\%$ for 8 receive antennas. These results are also an important indication that attackers with beam steering capability can hide in such scenarios including massive number of receive antennas. In other cases, the CNN architecture shows robust detection performance. Table \ref{table3} shows that the proposed lightweight CNN architecture is highly competent for detecting attacks in such scenarios for all SJRs, SNRs, jamming types, and receive antenna counts.
\begin{table}[!h]
\centering
\caption{Performance comparison of the proposed intelligent SDA detection for different numbers of receive antennas and all SNRs.}
\label{table3}
\adjustbox{width=0.75\columnwidth}{%
\begin{tabular}{|c|c|}
\hline
\textbf{Number of Receive Antennas} & \textbf{Accuracy (\%)} \\ \hline
4                                   & 99.95                  \\ \hline
8                                   & 98.33                  \\ \hline
\end{tabular}%
}
\end{table}
\section{Conclusions}
Satellite communications are especially critical for 6G telecommunications systems and the sustainability of large constellations with diverse missions. However, the new space offers many incentives, but also important challenges and risks such as collisions and information leakage. Therefore, we first clarified the motivations and challenges as well as the potential threats to satellite communications. Considering these aspects, we proposed an ISDAC system that creates awareness of the entire satellite environment through its autonomy, flexibility, and intelligence. It is provident considering the limited resources in space. In particular, we considered jamming attacks, which are common in satellite communications. We assume that the attacker is capable of behaving in different ways. To detect the presence or absence of the attacker, the proposed ISDAC system includes a lightweight CNN. The detection performance of the proposed system has been studied under 12 scenarios and the results demonstrate the superiority and robustness of the ISDAC system.
\bibliographystyle{IEEEtran}
\bibliography{ref}
\end{document}